\documentclass[final,3p,times]{elsarticle}
\biboptions{sort&compress}

\usepackage{hyperref}

\usepackage{graphicx}

\usepackage{amssymb}
\usepackage{amsmath}
\usepackage{upgreek}
\usepackage{lineno}
\usepackage{gensymb}
\usepackage[utf8]{inputenc}
\usepackage[dvipsnames]{xcolor}

\usepackage{fancyhdr}
\usepackage[normalem]{ulem}

\journal{Ultramicroscopy}

\begin{document}
\Large
 \color{red}
This is the Preprint version of our article send to Ultramicroscopy for publication.
\color{black}
\normalsize

\begin{frontmatter}

\title{Aberration-corrected transmission electron microscopy with Zernike phase plates}

\author[label1,label2]{Simon Hettler\corref{cor1}}
\author[label1,label2,label3]{Raul Arenal}
\address[label1]{Laboratorio de Microscopías Avanzadas (LMA), Universidad de Zaragoza, Zaragoza, Spain}
\address[label2]{Instituto de Nanociencia y Materiales de Aragón (INMA), Universidad de Zaragoza, Zaragoza, Spain}
\address[label3]{ARAID Foundation, Zaragoza, Spain}
\cortext[cor1]{hettler@unizar.es}

\begin{abstract}
We explore the possibility of applying physical phase plates (PPs) in combination with aberration-corrected transmission electron microscopy. Phase-contrast transfer characteristics are calculated and compared for a thin-film based Zernike PP, a hole-free (HF) or Volta PP and an electrostatic Zach PP, considering their phase-shifting properties in combination with partial spatial coherence. The effect of  slightly converging illumination conditions, often used in high-resolution applications, on imaging with PPs is discussed. Experiments with an unheated Zernike PP applied to various nanomaterial specimens and a qualitative analysis clearly demonstrates the general compatibility of PPs and aberration-corrected transmission electron microscopy. Calculations and experiments show the benefits of the approach, among which is a strong phase-contrast enhancement of a large range of spatial frequencies. This allows the simultaneous imaging of atomic-resolution structures and morphological features at the nanometer scale, with maximum phase contrast. The calculations can explain why the HFPP damps contrast transfer at higher spatial frequencies.
\end{abstract}

\end{frontmatter}

\section{Introduction}

Physical phase plates (PPs) for contrast enhancement in transmission electron microscopy (TEM) have been predominantly applied in biological and medical sciences where specimens typically offer low contrast \citep{PPReviewMalacHettler}. By induction of a relative phase shift between scattered and unscattered part of the electron wave in the back focal plane (BFP), PPs enhance phase contrast, which otherwise requires the introduction of aberrations such as defocus. In combination with aberration correction, PPs offer the possibility to strongly enhance phase contrast in high-resolution (HR)TEM images \citep{Gamm.2008EffectPP}. They potentially allow the acquisition of images made of pure phase contrast and the reconstruction of the object-wave function \citep{Gamm.2010} and thus may be an alternative to related holographic techniques in TEM \citep{Koch.2010}.

Numerous approaches to experimentally realize a PP have been proposed \citep{PPReviewMalacHettler}. When thinking of an application in HRTEM, PPs offering the induction of a variable phase shift combined with a minimum of matter introduced in the BFP seems the most promising approach. Such PPs allow to tune the contrast while keeping undesired obstruction or scattering to a minimum. Electrostatic PPs, which induce the phase shift by a localized electrostatic field at the position of the zero-order beam (ZOB) of unscattered electrons such as the Börsch \citep{BoerschOptimize.2007} and the Zach PP \citep{PPZach_MM2010} have already been applied on material science samples in HRTEM \citep{ZPP_Boersch_UM_2010,SimonZachPP_2016}. However, they could not reveal the full potential due to limitations of the PP design \citep{ZPP_Boersch_UM_2010} and of the employed microscope \citep{SimonZachPP_2016}. Although better suited, these PP types require specific holders with electrical feedthrough, which poses technological challenges especially in aberration-corrected microscopes with a small pole-piece gap. A potentially matter-free PP is the laser PP, which however requires large hardware modifications \citep{Glaeser_LaserPP_2019}.  In contrast, the most common and straightforward PP type is the Zernike PP \citep{NagayamaDanev2001b}, which consists of a thin amorphous carbon film with a circular hole that exploits the mean inner potential of amorphous matter to induce a fixed phase shift depending on the film thickness. This type of PP therefore does not require large modification of microscope hardware but comes at the cost of the loss of information in the PP due to scattering in the PP material itself. 
The hole-free (HF) or Volta PP \citep{HFPP_2012,VPP_2014}, which has had quite some success recently \citep{Kotani_Mag_2018,VPP_ApoferritinStruct_2019}, seems not ideally suited as the phase shift cannot be directly controlled \citep{2018_Simon_NegCharge}, the PP requires a heating element and has recently been reported to damp phase-contrast transfer at high spatial frequencies \citep{VPP_DQE_2020}.

In this work, we explore the possibility of combining an aberration-corrected microscope with a PP in a practical and theoretical way. We present experimental results using unheated Zernike PPs with varying cut-on frequencies applied on different nanomaterials, e.g. Fe\textsubscript{3-$\updelta$}O\textsubscript{4} magnetite nanoparticles (NPs) \citep{FeONPSartori.2021}. The effect of beam parallelity (convergence) is studied. By calculation of realistic phase-contrast transfer functions when imaging with PPs, we also give an explanation why the HF or Volta PP damps the contrast transfer at high spatial frequencies due to its sharp phase-shift profile.

\section{Materials and Methods}
\label{S:PracTheo}


An image-corrected Titan\textsuperscript{3} (Thermo Fisher Scientific) operated at 300 and 80~kV has been used to conduct the experiments. To insert a Zernike PP in the microscope, we designed a custom lamella for objective apertures (fabricated by Günther Frey GmbH, Berlin, Germany), which carries, in addition to objective apertures, a circular opening to implement conventional TEM grids (see Figure S1 in supplementary information (SI)). Zernike PPs made of amorphous C (aC) thin films with thicknesses of 27~nm and 17~nm were produced by electron-beam evaporation of graphite on mica substrates and subsequent floating on 200 mesh copper grids. Thicknesses were first measured by a profilometer and confirmed via TEM analysis of cross sections and correspond to a phase shift of $\approx\uppi$/2 at 300 and 80~kV assuming a mean inner potential of C of 9 V \citep{Dries.2016}. Experiments showed a phase shift of 0.45$\uppi$ and 0.56$\uppi$ for the PPs at 300 and 80~kV, respectively. A Helios 600 (Thermo Fisher Scientific) dual-beam instrument was employed to mill circular openings of different sizes (0.5 - 6~$\upmu$m radius) in the aC film. aC thin films with thicknesses smaller than 7~nm were used as samples and specimen support to study the phase-shifting properties of the Zernike PP.

\begin{figure}[t]
    \centering
    \includegraphics[width=0.78\linewidth]{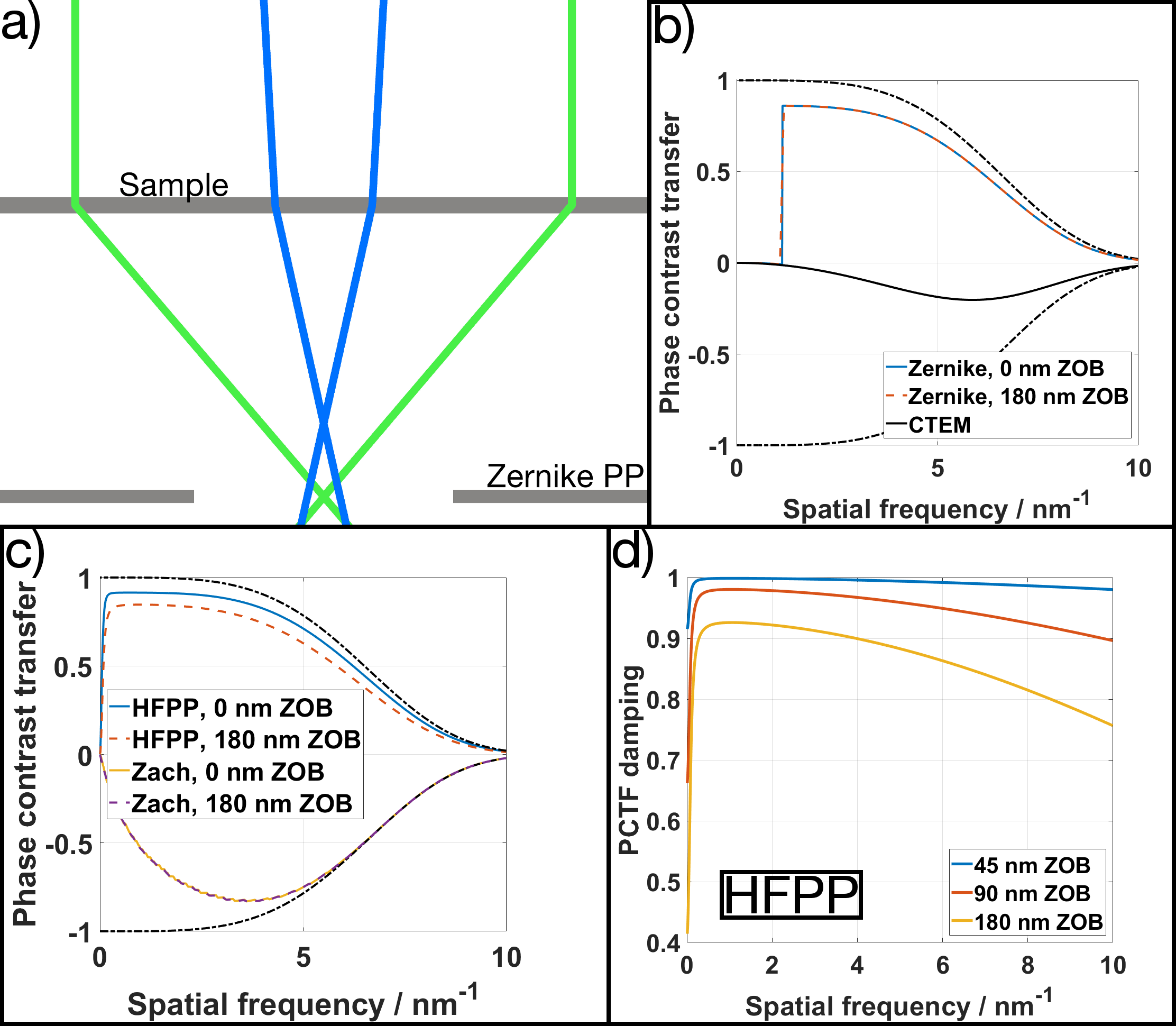}
    \caption{(a) Sketch of crossover position with respect to PP plane under parallel (green lines) and  coherent  convergent (blue) illumination. (b,c) Plots of PCTF for an aberration-corrected microscope showing CTEM at Scherzer defocus of -1.6~nm (black solid line) and different PPs considering various sizes of the ZOB in the PP plane caused by a finite source size (partial spatial coherence). Envelope function caused by partial coherence is plotted by dash-dotted lines. (b) PCTFs for Zernike PP with a hole with 4~$\upmu$m radius  with (dashed orange line) and without (solid blue line) consideration of a finite ZOB ($\alpha$~=~0.1~mrad). (c) PCTFs for Zach PP with a distance of 3~$\upmu$m between tip and ZOB  with (dashed purple line) and without (solid yellow line) consideration of a finite ZOB ($\alpha$~=~0.1~mrad) and the HFPP (width of phase-shift distribution of 500~nm)  with (dashed orange line) and without (solid blue line) consideration of a finite ZOB ($\alpha$~=~0.1~mrad). (d) Damping of PCTF amplitude due to finite ZOB sizes compared to a HFPP with an infinitesimally small ZOB  caused by varying phase shift within the ZOB.  }
    \label{F:SketchPCTF}
\end{figure}

The microscope was operated in TEM mode (spot size 3) using a 50~$\upmu$m C2 condenser aperture for the experiments at 300~kV and a 100~$\upmu$m C2 aperture at 80~kV. If not explicitly noted, experimental results were obtained at 300~kV. The condenser lens system with three lenses allows the parallel illumination of the sample for a range of illuminated sample areas. An adjustment of the condenser lens system was performed in addition to typical adjustment procedures before each microscopy session to reduce the movement of the ZOB in the BFP to a minimum when adjusting the image intensity (zooming). After implementation of the custom PP holder, the lenses were adjusted in order that the BFP coincides with the PP plane (green lines in Figure~\ref{F:SketchPCTF}a). To center the PP, we first spread the illumination strongly to visualize the PP hole on the screen at lower magnification and centered it using the conventional mechanics control of the aperture holder only. Fine centering was performed at higher magnification using the power spectrum.

The aberration corrector was tuned prior to inserting the PP. In principle, the acquisition of a Zemlin tableau \citep{Zemlin.1978}, necessary to correct aberrations, is also possible with a PP. A Zemlin tableau is acquired by deliberate beam tilting, which entails a movement of the ZOB in the PP plane and thus a change of the phase-shifting distribution of the PP for each image of the tableau. Correction methods would have to be adjusted taking into account these changes. However, in the case of an unheated Zernike PP, a negative charge builds up for each image of the tableau (see section~\ref{S:Charge}), which hinders their correct interpretation and introduces severe artifacts in the application of the PP.

For conventional HRTEM imaging, the intensity of the illumination in parallel mode is typically insufficient, requiring a further focusing of the beam on a smaller sample area by using a slightly convergent beam. This, however, moves the crossover position out of the BFP as indicated by the blue lines in Figure~\ref{F:SketchPCTF}a. This leads to a larger size of the ZOB in the PP plane, which, under the used conditions, amounts up to 2.5~$\upmu$m in diameter for a rather strongly condensed beam at a convergence angle of 1.4 mrad (see calibration Table S1 in SI).

A custom Matlab software was used to analyze the acquired images. Defocus, astigmatism and phase shift were determined by a pattern recognition process of power spectra described in \citep{Het2015}. Lower limit of defocus determination was approximately $\pm$50~nm. The central area of the power spectrum containing the Zernike PP hole was excluded from the recognition process as it is not phase shifted. The spatial phase-shift distribution in HFPP mode was obtained by analysis of the positions of the Thon-rings in an image series, similar as conducted for Figure~6 in \citep{HFPP_Pol2017}.

Electron-energy loss spectroscopy (EELS) was performed in TEM mode (image plane entering the filter) without specimen with a Gatan Image Filter Tridiem controlled by a custom Digital Micrograph (Gatan, Inc.) script.

\newpage

\section{Phase-contrast transfer function with PP}

To calculate the phase-contrast transfer function (PCTF) in presence of a PP, several effects have to be considered, which depend on the type of PP. They are discussed in detail in our recent simulation study \citep{PPSimsHettler.2021}. The PCTF for conventional (C)TEM  without PP is given by $sin(\chi(u))$ with the wave aberration function $\chi$ containing contributions from lens aberrations in dependence of the spatial frequency $u$ \citep{ReimerTEM_5}.  For a partially coherent source, the image is formed by the superposition of fully self-coherent single-electron wave functions, which each have slightly different energies (temporal) and source points (spatial). Field-emission guns have a high coherence allowing to describe the damping by partial coherence by envelope functions. Temporal coherence can as well be described by an incoherent sum using a weighted-focal series. The amount of partial temporal and spatial coherence is typically reflected by the microscope parameters focal spread $\Delta$ and semi-convergence angle $\alpha$.  

In presence of a PP, the phase-shift distribution $\phi_{PP}(u)$ can be added to $\chi(u)$. However, a simple addition of $\phi_{PP}(u)$ holds only true if the ZOB has an infinitesimally small diameter in the plane of the PP. This is a wrong assumption even under parallel illumination, where the ZOB possesses a finite diameter due to the finite source size (finite illumination aperture). For a correct consideration, one can follow the concept of a partially-coherent source, which is implemented by an incoherent sum over the finite source size. In combination with a PP, the finite source size can be translated to a shift in the PP plane.  Therefore, $\phi_{PP}(u)$ cannot be simply considered by a static term in $\chi$, because electrons stemming from slightly different points in the source have different incident angles and   pass through different points in the plane of the PP and "see" shifted  phase-shift distributions  $\phi_{PP}(u+\delta \xi)$. This effect has already been considered in image simulations  \citep{PPZach_MM2010,PPSimsHettler.2021}.  Here, we transfer this concept to  calculate a more realistic PCTF by a weighted sum over the finite ZOB size ($\delta \xi$) (corresponding to a finite source size and corresponding to different incident  angles) using a 2D Gaussian distribution (more details on calculation in SI, section SI1).

Figure~\ref{F:SketchPCTF}b and c show PCTFs for Zernike, HFPP and Zach PPs in dependence of the ZOB size and in comparison to conventional (C)TEM under Scherzer conditions. For the calculations conducted at 300~kV, a focal spread of $\Delta$~=~4.5~nm, a focal length f~=~1.8~mm and a spherical aberration constant of 1~$\upmu$m were assumed and higher-order aberrations were neglected. The semi-convergence angle determining partial spatial coherence was set to different values of $\alpha$=0, 0.025, 0.05 and 0.1~mrad, which translates into ZOB sizes of 0, 45, 90 and 180~nm. At the used defocus values (see below), the envelope for partial spatial coherence is negligible compared to temporal coherence for all of the values for $\alpha$.  The phase-shift distributions of the PPs were considered with a thickness corresponding to a phase shift of $\uppi$/2 for the Zernike PP and with $\pm\uppi$/2 at the central position of the ZOB for HFPP (-) and Zach PP (+) \citep{PPSimsHettler.2021}.  Scattering in the thin-film PPs was considered by applying a damping factor $A_{PP}=exp(-t/\lambda_{MFP})$ with a value of the mean free path (MFP) $\lambda_{MFP}$~=~180~nm and a mean inner potential of amorphous carbon of 9~V \citep{PPSimsHettler.2021}.  Defocus values of -1~nm for the PP PCTFs and -1.6~nm for the CTEM PCTFs (Scherzer defocus) were selected.

The PCTF for the Zernike PP when neglecting the finite ZOB size  (blue line in Figure~\ref{F:SketchPCTF}b) first follows the CTEM transfer until it reaches the edge of the hole (4~$\upmu$m radius) where it jumps to a value of roughly 0.86, which is smaller than the possible maximum due to damping caused by scattering in the Zernike thin film. When considering a finite ZOB size, the only effect is a smoothing of the hard step at the hole edge (dashed orange line in Figure~\ref{F:SketchPCTF}b). 

It is worth noting, how the electrons scattered in the Zernike PP thin film contribute to the final image contrast. If the illuminated area of the specimen is not substantially larger than the area imaged by the camera, basically no electron scattered in the Zernike thin film reaches the camera. The Zernike PP thus partially acts as a small objective aperture as these scattered electrons add a constant bright-field type contrast to the image. This contrast has the same sign as the positive phase contrast caused by the Zernike PP.

In case of the Zach PP, barely no effect for a finite ZOB size  is observed if the Zach PP is positioned rather far away from the ZOB (distance between PP tip and ZOB of 3~$\upmu$m) as the potential distribution at the ZOB position is almost flat (Figure~\ref{F:SketchPCTF}c). This is in contrast to the HFPP, which has a sharp phase-shift distribution closely matched to the ZOB size. A width of the phase-shift distribution of 500~nm and a film thickness of 8~nm is assumed for the plots displayed in Figure~\ref{F:SketchPCTF}c and a clear reduction of the PCTF with an increase in ZOB size is observed. The reduction of the PCTF for the HFPP when considering the finite ZOB size  is shown in dependence of the spatial frequency in Figure~\ref{F:SketchPCTF}d. A rather strong reduction is observed for already small ZOB sizes and the reduction is additionally dependent on the spatial frequency, showing a further decrease at higher frequencies.  The chosen ZOB sizes roughly correspond to a tenth, fifth and third of the used size of $\phi_{HFPP}$ of 500 nm. Experiments have shown that the size of $\phi_{HFPP}$ is closely matched to the ZOB creating the phase-shifting patch and follows a Gaussian+Lorentzian shaped profile, which has a sharp central peak (Gauss) and longer-ranging tails (Lorentz) \citep{2018_Simon_NegCharge,Pretzsch.2019}. Therefore it seems a valid assumption that $\phi_{HFPP}$ is varying strongly around u=0, making the choice of ZOB sizes in combination with the width of $\phi_{HFPP}$ for Figure~\ref{F:SketchPCTF}d a realistic one. In sum,  this result, obtained by the assumption of a finite size of the ZOB caused by partial spatial coherence leads to an additional envelope for the HFPP. This  can very well explain the experimentally observed damping of the Volta PP. This damping was observed to be higher than expected only from scattering in the film and additionally further decreases with increasing spatial frequencies corresponding well with Figure~\ref{F:SketchPCTF}d  \citep{VPP_2014,VPP_DQE_2020}. This finding is not only true for the Volta or HFPP, but for any PP that exhibits a non-flat phase shift distribution at the ZOB position , e.g., also for the laser PP \citep{Glaeser_LaserPP_2019}. 

When comparing the PP-PCTFs with CTEM, it becomes evident that the total phase contrast, which is given by the area under the curves, is drastically increased when using a PP \citep{Gamm.2008EffectPP}. The transfer with PP is not only higher at lower spatial frequencies but also excels the transfer at higher frequencies reached by CTEM under Scherzer conditions. In case of the Zernike PP, the PCTF under consideration of a finite ZOB size seems favorable as the hard edge of the PP, typically causing fringing artifacts, is smoothed, which can reduce these artifacts \citep{ZPPOptimizing_NagayamaDanev_UM2011,ObermairGradedZPP}. In the recent study of graded Zernike PPs, the consideration of a finite ZOB size yielded a better match between simulation and experiment \citep{ObermairGradedZPP}. 

When leaving parallel-beam conditions to increase intensity in the investigated area by deliberately introducing a coherent convergence angle, the crossover position leaves the plane of the PP (Figure~\ref{F:SketchPCTF}a). In contrast to the effect of partial coherence, this coherent convergence angle is an intrinsic property of the single-electron wave function and leads to a varying incident angle of the wave within the illuminated sample area, being flat in the center and increasing towards the edge of the illuminated area. This can be observed in experiments by calculating power spectra of quadrants of an acquired image, which differ from the power spectrum of the whole image in case of a coherent converging illumination. In combination with a PP this results in a phase-shift distribution $\phi_{PP}(u)$ that is dependent on the position within the image. 

As a HFPP requires the ZOB to be focused on the HFPP plane, it is not well suited to be used under coherent convergent-beam conditions. The amount of phase shift and the size of the phase-shift distribution $\phi_{PP}(u)$ will be strongly affected, making it not suitable for HRTEM applications. In contrast, both Zernike and Zach PP can be expected to tolerate moderate convergent-beam conditions as both show a rather flat phase-shift distribution around u=0.  We experimentally studied imaging under coherent converging conditions typical for HRTEM by acquisition of a series of Zernike PP TEM images using varying coherent  convergence angles (Figure~S2 in SI). The analysis of images and power spectra shows that it is possible without restraints if the hole of the Zernike PP is large enough. The coherent converging beam results in a slightly varying cut-on frequency within the image for the Zernike PP, which led to no observable effect in the experimental images.

\newpage
\begin{figure}[t]
    \centering
    \includegraphics[width=0.83\linewidth]{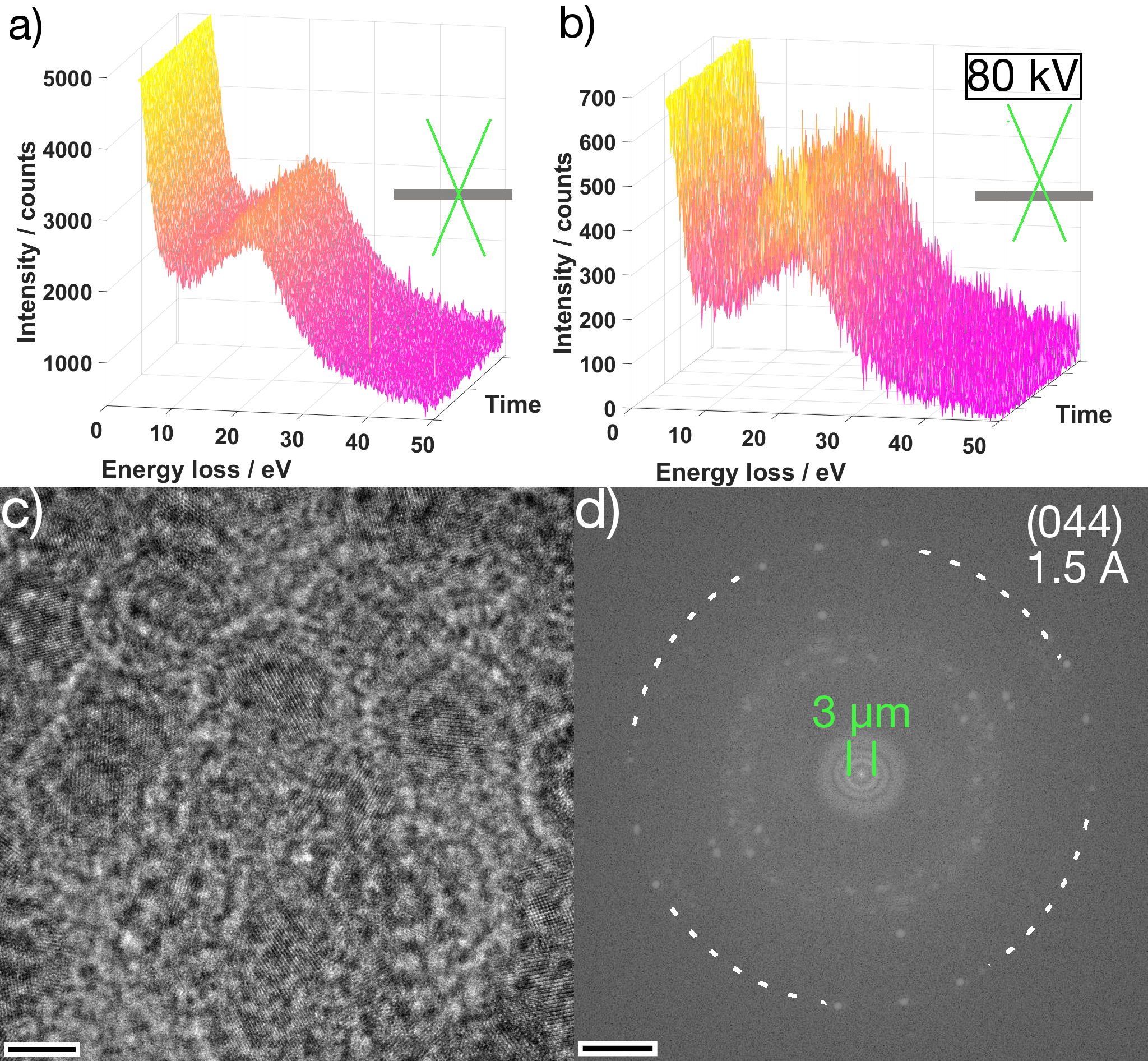}
    \caption{(a,b) Evolution of EEL spectra after insertion of the Zernike PP at a position without hole. (a) No change in inelastically scattered intensity is observed with the ZOB focused on the PP. (b) A small initial decrease in inelastically scattered intensity is seen at 80~kV with the beam spread over several $\upmu$m on the PP. (c) Aberration-corrected TEM image of Fe\textsubscript{3-$\updelta$}O\textsubscript{4} nanoparticles on an aC support film acquired with Zernike PP with a smaller hole radius of 1.5 $\upmu$m (cut-on frequency of $\approx$ 2~nm). (d) The power spectrum reveals the charged PP (ringing around the hole of the PP) as well as reflections corresponding to different distances of the Fe\textsubscript{3-$\updelta$}O\textsubscript{4} magnetite spinel structure. The (044) distance at 0.15 nm is marked by a white circle. Scale bars are (c) 5~nm and (d) 2~nm\textsuperscript{-1}. }
    \label{F:Smallcuton}
\end{figure}

\section{Charging behaviour and stability of the unheated Zernike PP}
\label{S:Charge}

Contamination and charging of PPs can be a severe problem that may introduce image artifacts. Although we did not heat the Zernike PP, contamination is not observed as proven by the acquisition of electron energy-loss (EEL) spectra with the ZOB focused on the Zernike film (Figure~\ref{F:Smallcuton}a). The low-loss EEL spectra are constant and no growing inelastic signal is seen, which would be caused by a deposition of contamination \citep{2017_Simon_Contami_Micron}. In contrast, in the case of a ZOB not focused on the Zernike film, but spread over an area with several $\upmu$m in diameter, even a slight decrease of the inelastic signal is observed (Figure~\ref{F:Smallcuton}b). This decrease is attributed to the desorption of chemisorbed molecules, which is thought to be the cause for the negative charging of the Volta and HFPP \citep{2018_Simon_NegCharge}. The acquisition of TEM images in HFPP mode, i.e. with the ZOB focused on an area of the PP thin film without a hole, clearly reveals the buildup of negative charge (Figure S3 in SI), which can amount up to several $\uppi$ after a few seconds, similar to previous studies on unheated films \citep{Pretzsch.2019}.

The presence of negative charging under intense illumination poses a rather strong limitation to the application of unheated Zernike PPs in HRTEM as samples are typically crystalline and may cause strong reflections, which can generate charging on the film. Such a charge will add an additional phase shift to the electrons of the reflection spot and surrounding areas, making a correct interpretation of the resulting image almost impossible except for special cases \citep{KenGrating_2020}. In case of NP samples however, the intensity in the diffracted spots is small and a charging can be avoided for the time needed for an experiment. Figure S4 and S5 in the SI show two stability studies of Fe\textsubscript{3-$\updelta$}O\textsubscript{4} magnetite and MoS\textsubscript{2} NPs, where the lattice-fringe contrast in Zernike PP HRTEM images remains stable throughout the investigated time period indicating the absence of (significant) charging for such samples. In contrast, 2D materials spread over larger areas, such as graphene, cause much stronger reflections and therefore indeed create their proper charged areas (Figure~S6 in SI). 

Charging is as well a problem when working with Zernike PPs with relatively small holes where the intense ZOB is more likely to hit the Zernike PP close to the hole during adjustment or imaging. Figure~\ref{F:Smallcuton}c,d shows the application of a Zernike PP with a hole radius of 1.5~$\upmu$m on Fe\textsubscript{3-$\updelta$}O\textsubscript{4} magnetite NPs. While the lattice-fringe contrast of the NPs is revealed, charging of the film area close to the hole is visible in the power spectrum  at lower spatial frequency and causes artifacts in the image contrast complicating the image interpretation.

\begin{figure}[t]
    \centering
    \includegraphics[width=\linewidth]{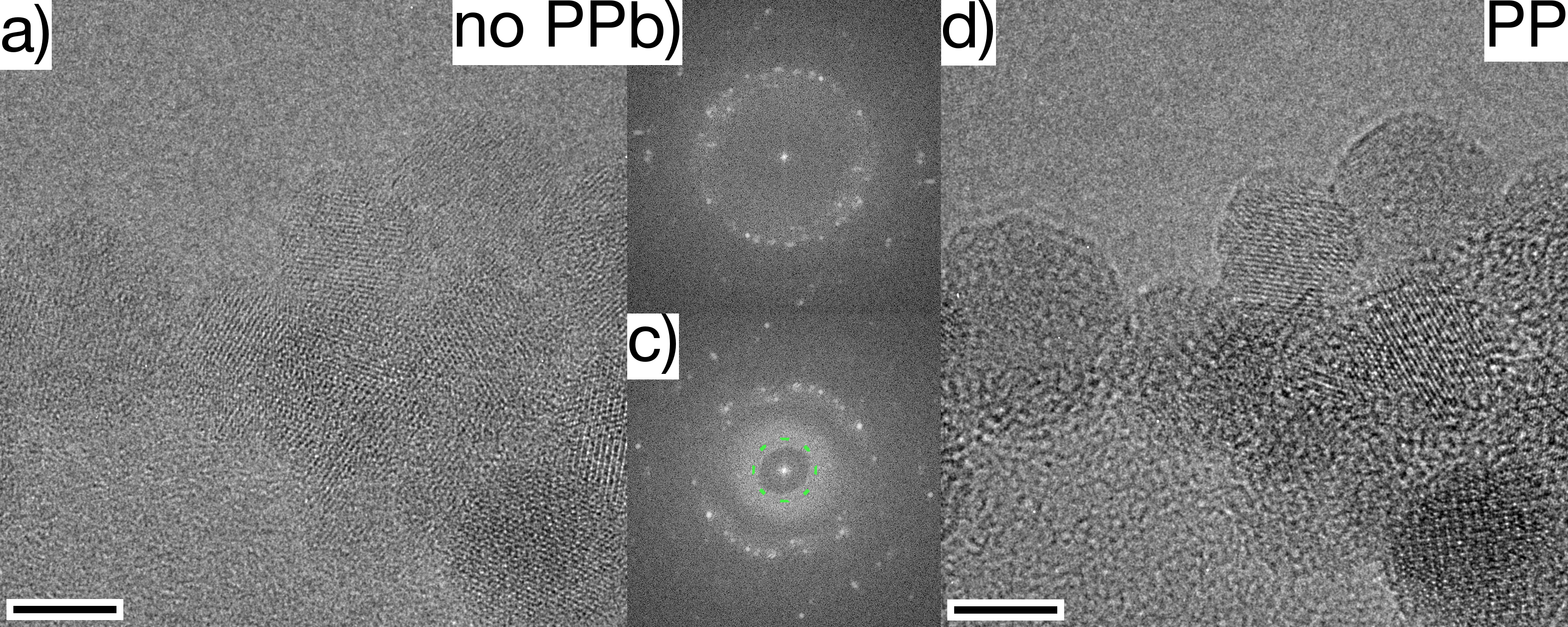}
    \caption{(a,d) HRTEM images of Fe\textsubscript{3-$\updelta$}O\textsubscript{4} magnetite NPs and (b,c) corresponding power spectra acquired  (a,b) without and (c,d) with Zernike PP (hole radius of 5~$\upmu$m). The images are displayed at identical contrast settings, mean image intensity is (a) 144.2 and (b) 141.8 counts. Convergence angle was approximately 1.2~mrad. Scale bars are (a,d) 5~nm and the power spectra have a size of 14.3 nm\textsuperscript{-1} x 14.3 nm\textsuperscript{-1}.}
    \label{F:PPnoPP}
\end{figure}

\begin{figure}[ht]
    \centering
    \includegraphics[width=0.82\linewidth]{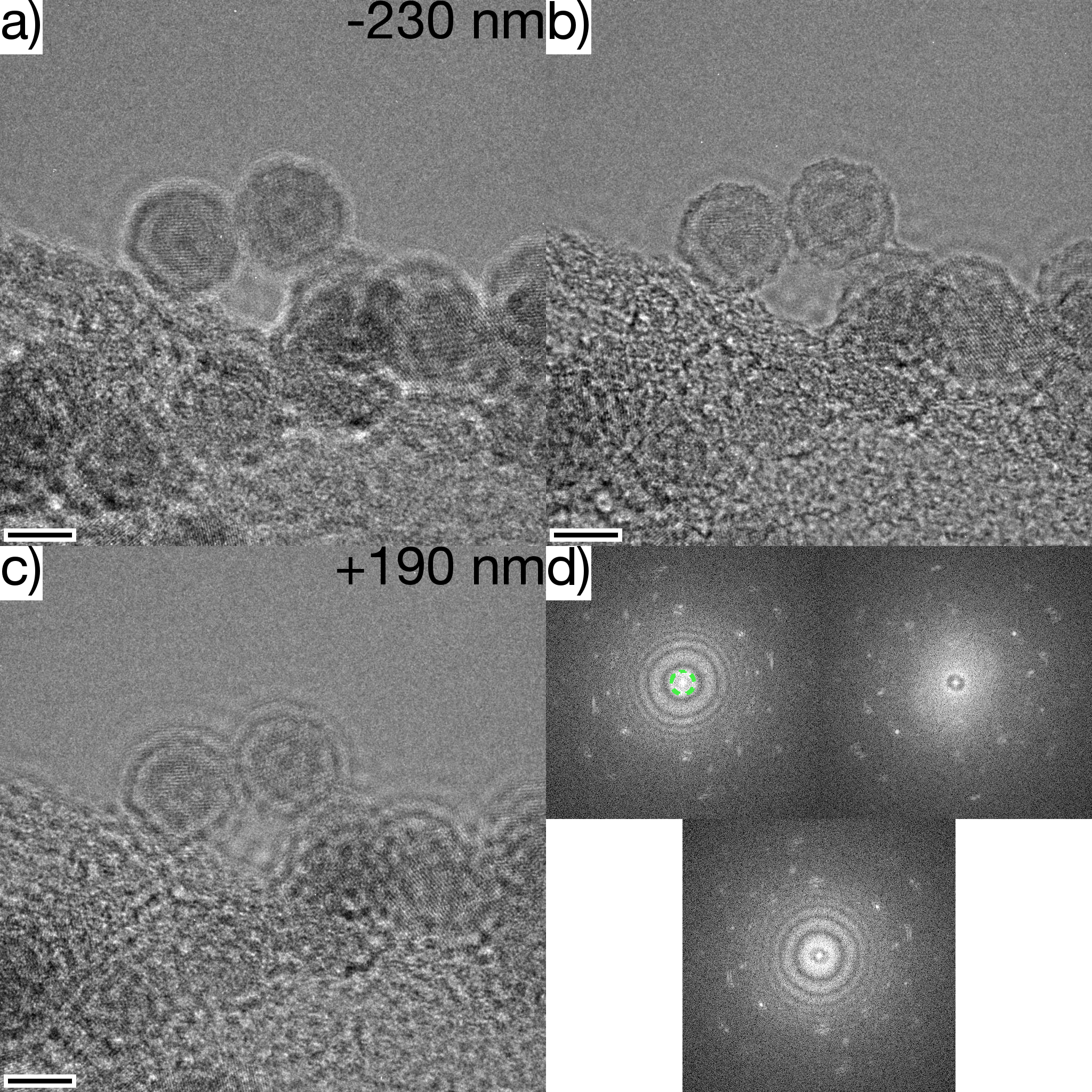}
    \caption{(a-c) HRTEM images of Fe\textsubscript{3-$\updelta$}O\textsubscript{4} magnetite NPs and (d) corresponding power spectra acquired with Zernike PP with a smaller hole radius of 2~$\upmu$m and a defocus value of (a) -230~nm, (b) $\approx$0~nm and (c) +190~nm. Scale bars are (a-c) 5~nm and the power spectra are displayed up to 7.1 nm\textsuperscript{-1}.}
    \label{F:MediumCuton}
\end{figure}

\section{Aberration-corrected HRTEM with Zernike PP}

Figure~\ref{F:PPnoPP}a shows a HRTEM image of Fe\textsubscript{3-$\updelta$}O\textsubscript{4} NPs on an aC support film close to Gaussian focus. Lattice fringes are visible in most NPs and the power spectrum reveals reflections rings at 0.3~nm (022), 0.25~nm (113), 0.16~nm (115) and 0.15~nm (044) corresponding to different lattice distances of the magnetite spinel crystal structure \citep{FeONPSartori.2021}. Figure~\ref{F:PPnoPP}d shows a HRTEM image of the identical sample area as in (a), which has been obtained directly after insertion of a previously centered Zernike PP, without changing any additional imaging parameter. The overall contrast of the NPs and the supporting film is considerably increased and also the lattice-fringe contrast in almost all NPs is increased. While the contrast at intermediate spatial frequencies is solely enhanced by the Zernike PP, the lattice-fringe contrast is affected as well by a possible minor change of defocus. The power spectra suggest a defocus change smaller than 25~nm, which however already would result in a significant modification in contrast of atomic-resolution images. There is no indication that the PP has a negative effect on resolution and a small defocus change may already be observed only by inserting an objective aperture. Comparison of defocused TEM images with and without PP show a defocus change of up to 30~nm, which can go in hand with a small change of astigmatism (Figures S7 and S8 at 80~kV). These changes are attributed to charged areas on the objective aperture holder and an increase in magnitude of the effect was observed after an extensive use of the PPs. This additional defocus and possible astigmatism can be adjusted and corrected as done conventionally.

\begin{figure}[t]
    \centering
    \includegraphics[width=0.9\linewidth]{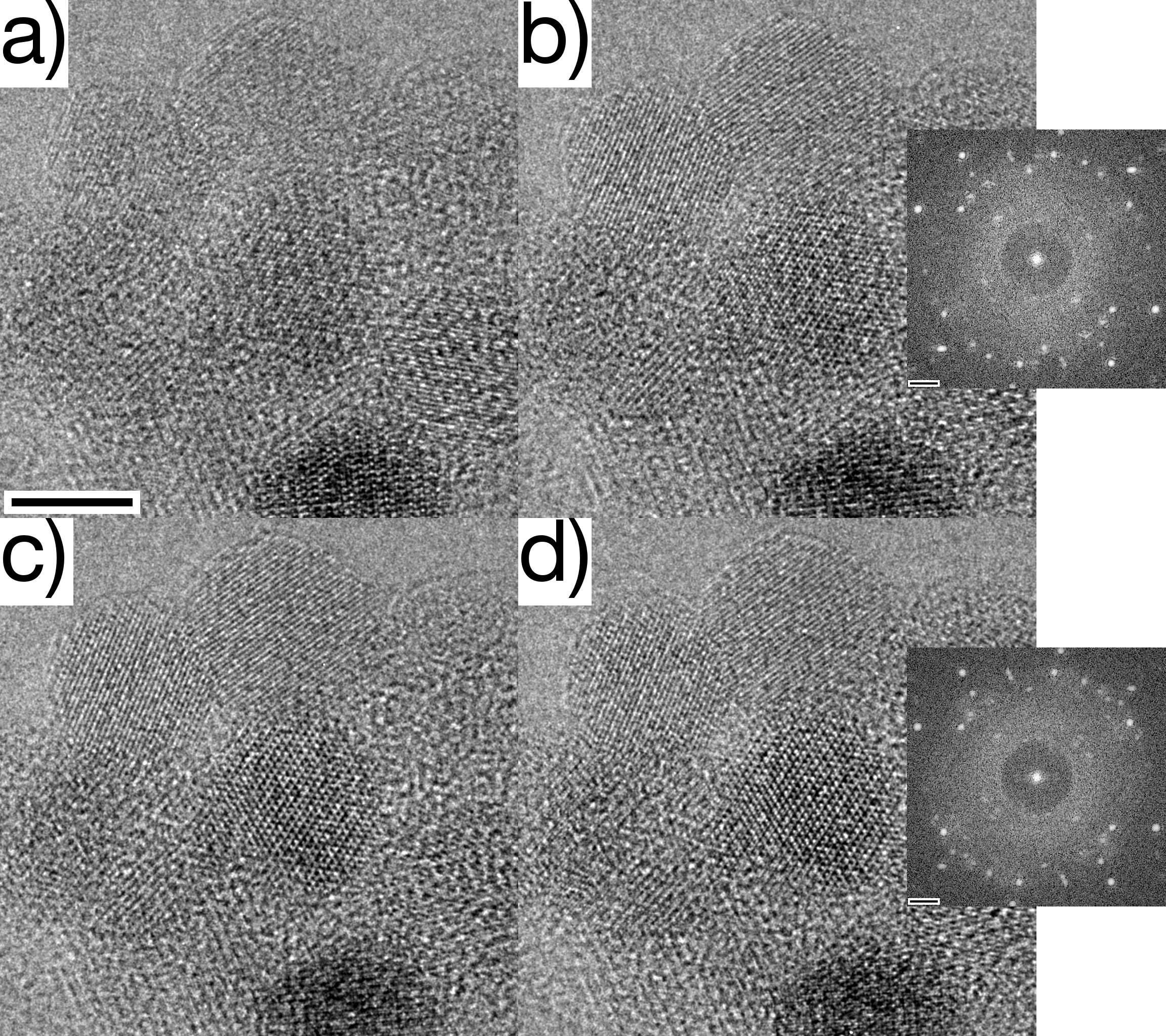}
    \caption{HRTEM images of Fe\textsubscript{3-$\updelta$}O\textsubscript{4} magnetite NPs at the edge of an aC film acquired with a Zernike PP (hole radius 5~$\upmu$m) close to Gaussian focus. Coherent  convergence angle was approximately 1.2~mrad. The defocus value was decreased by 5~nm for each image going from (a) to (d). In contrast to imaging without PP, the lattice-fringe contrast is strong and does not change sign around the Gaussian focus. Scale bar is 5~nm and 1~nm\textsuperscript{-1} for the inset power spectra. }
    \label{F:ConstantContrast}
\end{figure}
The image intensity when inserting the PP drops slightly from a mean value of 144.2 to 141.8 counts, which is less than 2\%. This drop is determined by both the number of electrons scattered in the sample and in the PP, as only electrons scattered in the specimen are actually transmitted through the PP film. The increase in contrat is clearly seen in the image and also manifests itself in the value of the full-width at  tenth maximum of the histogram, which increases by 14\% from 59 to 67. The increase is primarily observed to lower values, which is consistent with the positive (dark) phase contrast observed for Zernike PP images.

Due to the large hole with a radius of 5~$\upmu$m, contrast fringing at sample edges is almost absent in Figure~\ref{F:PPnoPP}d. Figure~\ref{F:MediumCuton} shows three HRTEM images acquired with a Zernike PP with a smaller hole radius of 2~$\upmu$m and different defocus values. In the image acquired close to Gaussian focus (b), contrast fringing at the sample edges and around the NPs can be observed. The aC thin film and the NPs again appear with increased overall contrast when compared to in-focus CTEM imaging. The images in Figure~\ref{F:MediumCuton}(a,c) acquired at over- (+190~nm) and underfocus (-230~nm) values, which are large values for HRTEM, show that the overall contrast is not further increased in a considerbale manner and the fringing at the edges remains similar. Both effects may therefore be mainly attributed to the action of the PP. However, delocalization effects become clearly visible as the lattice fringes are observed outside the actual NP and the image appears blurred. 

The big advantage of the application of PPs in HRTEM is that under focused conditions, phase contrast is maximum and does not change the sign as it does in CTEM when going from over- to underfocus \citep{PPSimsHettler.2021}. This effect is demonstrated by Figure~\ref{F:ConstantContrast}, which displays four TEM images acquired with Zernike PP at different defocus values around Gaussian focus. The difference in defocus between the individual images is 5~nm, going in direction of underfocus from (a) to (d). While changes in the lattice-fringe contrast in the NPs can be seen, the contrast does not change its sign. Position of focused conditions was confirmed by images acquired at larger over- and underfocus values.  We note that the requirement for focused conditions with PPs remains similar to CTEM and that the focusing procedure with PPs is largely different to CTEM as contrast is not minimum at Gaussian focus. 

\begin{figure}[ht]
    \centering
    \includegraphics[width=0.85\linewidth]{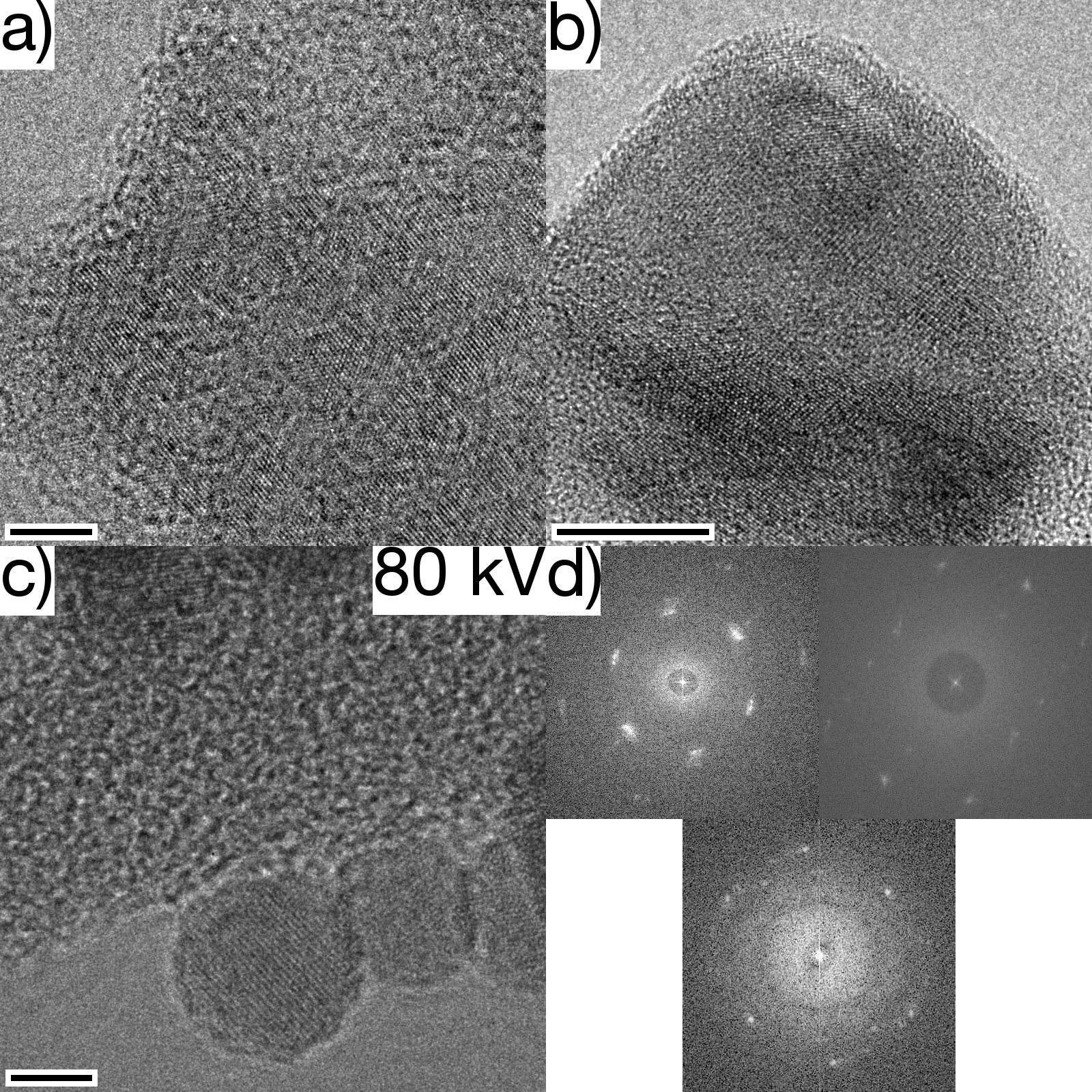}
    \caption{HRTEM images with Zernike PP of (a) W(S,Se)\textsubscript{2}, (b) MoS\textsubscript{2}-In\textsubscript{2}Se\textsubscript{3} and (c) Fe\textsubscript{3-$\updelta$}O\textsubscript{4} magnetite NPs at 80~kV. Hole radii of Zernike PP were (a) 3~$\upmu$m, (b) 5~$\upmu$m and (c) 6~$\upmu$m. Scale bars are (a) 5~nm, (b) 6~nm and (c) 4~nm. Power spectra are displayed up to 8.2, 5.6 and 5~nm\textsuperscript{-1}.}
    \label{F:AddApps}
\end{figure}

Figure~\ref{F:AddApps} shows the application of the Zernike PP on two additional nanomaterials. Figure~\ref{F:AddApps}a shows an image of a few-layer sheet of W(S,Se)\textsubscript{2} \citep{WSSe_Sreeda_2022}. The lattice is clearly resolved and in contrast to conventional HRTEM, the overall contrast is stronger, allowing the observation of the morphology and amorphous regions of the sheet. The six-fold set of reflections is clearly resolved in the corresponding power spectrum (Figure~\ref{F:AddApps}d). The second specimen is a small particle of a MoS\textsubscript{2}-In\textsubscript{2}Se\textsubscript{3} hybrid \citep{Mohapatra.2020}, where the lattice fringes are again clearly resolved. Figure~\ref{F:AddApps}c shows an image acquired at 80~kV of a Fe\textsubscript{3-$\updelta$}O\textsubscript{4} NPs sample. The resolution and sharpness of the image is reduced compared to the application at 300~kV, which is both attributed to the microscope and PP performance. At 80~kV, the damping caused by a PP inducing a phase shift of $\uppi$/2 is slightly stronger than at 300~kV and also charging of the PP is more readily observed.

\section{Summary and Outlook}

We studied the possibility of combining physical phase plates (PPs) with aberration-corrected transmission electron microscopy (TEM). The experimental results obtained with an unheated Zernike PP made of an amorphous carbon thin film with hole prove the general compatibility and its advantages. Nanomaterial specimens could be imaged without artifacts and the image contrast can be well explained qualitatively. The calculation of the phase-contrast transfer for PPs taking into account the finite zero-order beam (ZOB) size caused by partial spatial coherence  shows that the transfer is only marginally affected for Zernike and Zach PP.  Experiments show that the Zernike PP tolerates coherent converging-beam conditions typical for HRTEM imaging.

This work confirms that an application of PPs in aberration-corrected TEM in material science is highly promising. The additional possibilities can strongly improve imaging of weak-phase and weak-contrast nanomaterials and possibly allows their analysis in a more quantitative way. A heated, charging-free Zernike PP could improve phase contrast of many specimens in HRTEM applications without requiring large modifications of the microscope hardware.  Electrostatic PPs would offer a highly flexible tool to control and optimize phase contrast.  PPs with a sharp phase-shift profile that is varying within the ZOB, such as the hole-free or Volta PP \citep{HFPP_2012,VPP_2014} and the laser PP \citep{Glaeser_LaserPP_2019}, are well suited to improve phase contrast at low spatial frequencies. However, such a sharp phase-shift profile causes an additional damping envelope in combination with partial spatial coherence making these PPs not ideally suited for high-resolution applications.  

\section*{Acknowledgements}

The authors acknowledge funding by German Research Foundation (DFG project He 7675/1-1), from the European Union’s Horizon 2020 research and innovation programme under the Marie Sklodowska-Curie grant agreement No 889546, by the Spanish MICINN (PID2019-104739GB-100/AEI/10.13039/501100011033) and from the European Union H2020 programs “ESTEEM3” (Grant number 823717) and "Graphene Flagship" CORE 3 (Grant number 881603). The microscopy works have been conducted in the Laboratorio de Microscopias Avanzadas (LMA) at Universidad de Zaragoza. Support from R. Valero (LMA) with electron-beam evaporation of aC thin films is acknowledged. Sample courtesy from K. Sartori and B. Pichon (Fe\textsubscript{3-$\updelta$}O\textsubscript{4} NP, Université de Strasbourg, CNRS, France), MB Sreedhara and R. Tenne (W(S,Se)\textsubscript{2}, Weizmann Institute of Science, Israel) and A. Ismach (MoS\textsubscript{2}-In\textsubscript{2}Se\textsubscript{3}, Tel Aviv University, Israel) is acknowledged.

\bibliographystyle{citstyle.bst}
\bibliography{ZernikeHRTEM.bib}

\end{document}


\begin{frontmatter}

\title{Supplementary information: Aberration-corrected transmission electron microscopy with Zernike phase plates}

\author[label1,label2]{Simon Hettler\corref{cor1}}
\author[label1,label2,label3]{Raul Arenal}
\address[label1]{Laboratorio de Microscopías Avanzadas (LMA), Universidad de Zaragoza, Zaragoza, Spain}
\address[label2]{Instituto de Nanociencia y Materiales de Aragón (INMA), Universidad de Zaragoza, Zaragoza, Spain}
\address[label3]{ARAID Foundation, Zaragoza, Spain}
\cortext[cor1]{hettler@unizar.es}

\end{frontmatter}

\begin{figure}[h]
\centering
    \includegraphics[width=\linewidth]{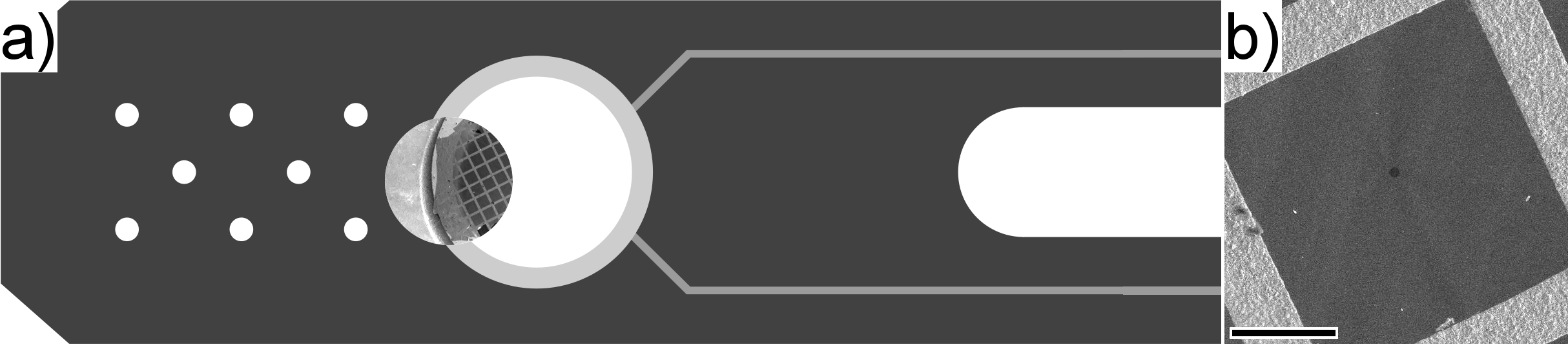}
    \caption{(a) Sketch of employed objective aperture lamella with inserted SEM image of mounted grid carrying an amorphous carbon thin film used as Zernike PP. (b) Enlarged SEM image of single hole of the grid with hole for Zernike PP in center. Scale bar is 50$\upmu$m.}
    \label{F:Fab}
\end{figure}

\newpage

\section{Calculation of a PCTF with PPs}

The PCTFs considering partial spatial coherence and a realistic phase-shift distribution of the PPs is calculated in dependence of the spatial frequency (u) by the following equation:

\begin{equation}
    PCTF(u) = \sum_{\delta \xi} B_{\delta \xi} exp\left[-\frac{\delta \xi_x^2}{2 \Xi ^2}-\frac{\delta \xi_y^2}{2 \Xi ^2}\right] \sum_{\delta Z} \frac{B_{\delta Z}}{2 \pi \Delta} exp\left[-\frac{\delta Z^2}{2 \Delta ^2}\right] \cdot sin\left[\chi(u,Z+\delta Z,\delta \xi)\right] \cdot E_S(u,Z+\delta Z) \cdot A_{PP}(u,\delta \xi)
    \label{Eq:PCTF}
\end{equation}

where the first sum (partial spatial coherence in combination with a PP) goes over the (x,y) position $\delta \xi$ in the ZOB with width $\Xi$ (identical in x and y direction) and with $B_{\delta \xi}$ the normalization factor in order that $\sum_{\delta \xi} B_{\delta \xi} exp\left[-\frac{\delta \xi_x^2}{2 \Xi ^2}-\frac{\delta \xi_y^2}{2 \Xi ^2}\right]=1$. The second sum goes over the defocus variation $\delta Z$ and calculates the effects of partial temporal coherence.  $B_{\delta Z}$ is the normalization factor for the second sum in order that $\sum_{\delta Z} B / (2 \pi \Delta ) \cdot exp\left[-\delta Z^2 /(2 \Delta ^2)\right] = 1$ with the focal spread $\Delta$. The wave aberration function $\chi(u,Z+\delta Z,\delta \xi)$ is given by:

\begin{equation}
    \chi(u,Z+\delta Z,\delta \xi) = \pi (Z+\delta Z) \lambda u^2 + \frac{\pi}{2} C_S \lambda^3 u^4 + \phi_{PP} (u,\delta \xi) + ...
    \label{Eq:WaveAb}
\end{equation}

with the electron wavelength $\lambda$, defocus $Z$, constant of spherical aberration $C_S$ and the phase-shift distribution of the PP $\phi_{PP}$, which is calculated in dependence of u for every position in the ZOB $\delta \xi$. The envelope considering spatial coherence with the semi-convergence angle $\alpha$ is given by 

\begin{equation}
    E_S (\vec{u})= exp \left( - \frac{\alpha^2}{4 \lambda^2} \left(\frac{\partial \chi (\vec{u})}{\partial \vec{u}}\right)^2 \right)
    \label{Eq:EnvelSpatialCoherence}
\end{equation}

We neglected the dependence of $E_S$ on the position within the ZOB $\delta \xi$ (corresponding to different incident angles) as for the studied defocus values, $E_S$ is negligible in comparison with the effect of temporal coherence.

\newpage

\section{Beam convergence}

\begin{figure}[h]
\centering
    \includegraphics[width=0.9\linewidth]{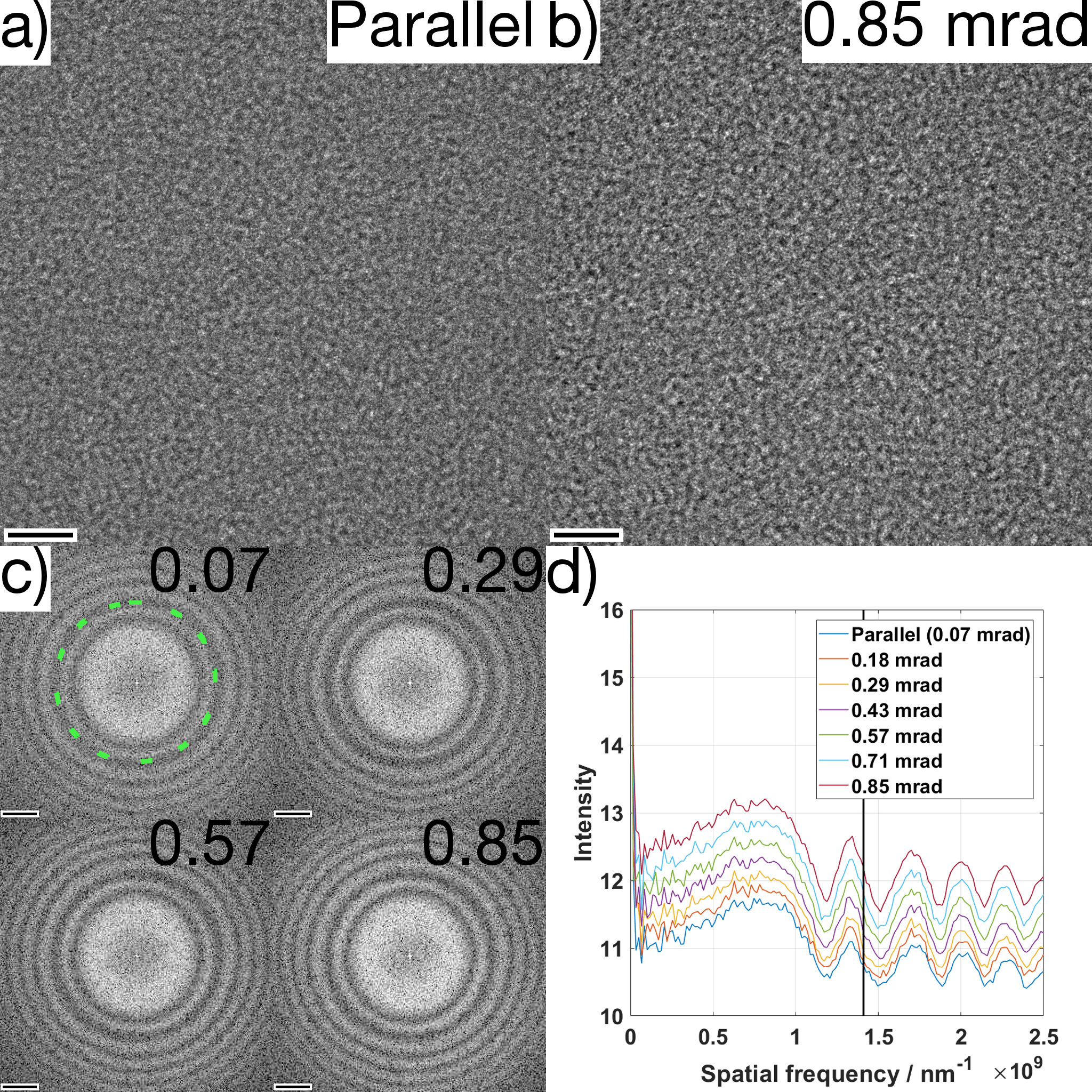}
    \caption{(a,b) Two TEM images of an amorphous C thin film acquired with Zernike phase plate (5 $\upmu$m hole radius) at different coherent convergence angles of (a) 0.07 and (b) 0.85 mrad. (c) Comparison of power spectra at different coherent convergence angles reveals similar appearance of Thon rings. Zernike PP hole edge is indicated by dashed green circle. (d) Rotationally averaged line profiles of power spectra show an increasing intensity and a minor change at the position of the PP edge marked by the black vertical line. Scale bar is (a,b) 7 nm and (c) 0.6 nm\textsuperscript{-1}. }
    \label{F:IAesr}
\end{figure}

\newpage

\begin{table}[h]
    \centering
    \begin{tabular}{|c|c|c|c|}
    \hline
\textit{Iluminated area} / nm & ZOB size / nm\textsuperscript{-1} & ZOB size / mrad & ZOB size / nm in BFP \\
    \hline
624 (parallel) &	0.03 &	0.07  &	130 \\
520 &	0.15 &	0.29  &	514 \\
420 &	0.29 &	0.56 &	1020 \\
320 &	0.43 &	0.85 &	1530 \\
220 &	0.57 &	1.12  &	2020 \\
120 &	0.72 &	1.41 &	2540 \\
\hline
    \end{tabular}
    \caption{Calibration of zero-order beam size in back focal plane (coinciding with PP plane) for different \textit{Illuminated areas} for a 50 $\upmu$m Condenser (C2) aperture ranging from last parallel setting (624 nm) to coherent converging illumination. Data is obtained by acquisition of diffraction patterns without specimen inserted and measuring the size of the ZOB. The size for parallel illumination is probably overestimated due to the high current density and a saturation of the camera.}
    \label{tab:IACalib}
\end{table}

\newpage

\section{Stability of unheated Zernike PP}

\begin{figure}[h]
\centering
    \includegraphics[width=0.95\linewidth]{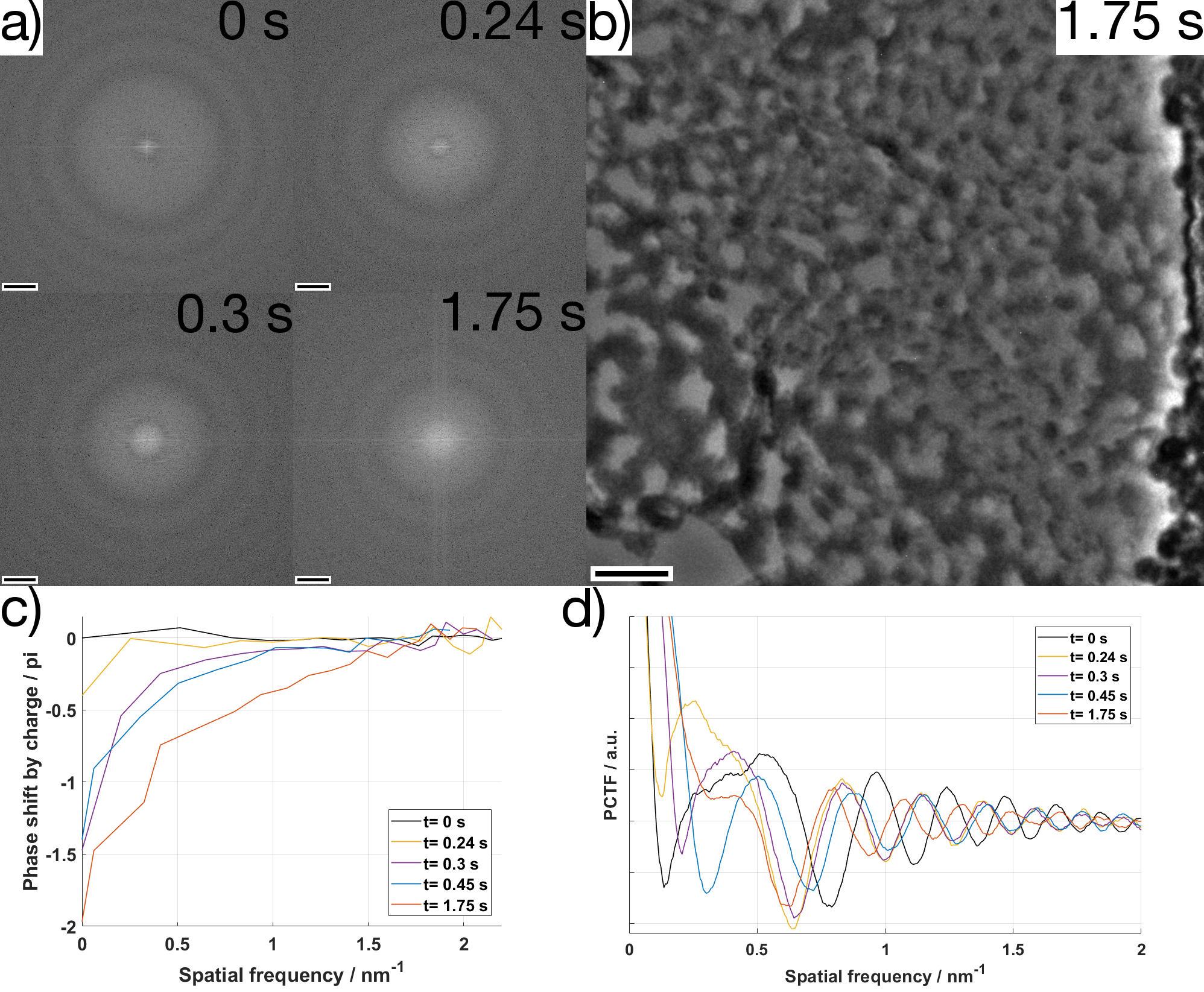}
    \caption{(a) Four power spectra of TEM images of a very thin amorphous C thin film acquired before (0s) and after insertion of the (unheated) Zernike PP thin film with no hole in the center. The strong buildup of the charge/phase shift is already evident in the first power spectra after insertion (0.24s) and further increases during the series. (b) TEM image at the end of the series shows strong contrast. (c) Using the Thon-ring positions (see d) of each power spectrum allows to detect the spatial distribution of the charge/phase shift build up on the PP. The negative phase shift increases in amplitude and also in width. (d) Plot of rotationally averaged profiles of the power spectra shows the strong change of the phase-contrast transfer, especially at lower spatial frequencies. Scale bars are (a) 0.3 nm\textsuperscript{-1} and (b) 20 nm. }
    \label{F:HFPP}
\end{figure}
\newpage

\begin{figure}[h]
\centering
    \includegraphics[width=0.95\linewidth]{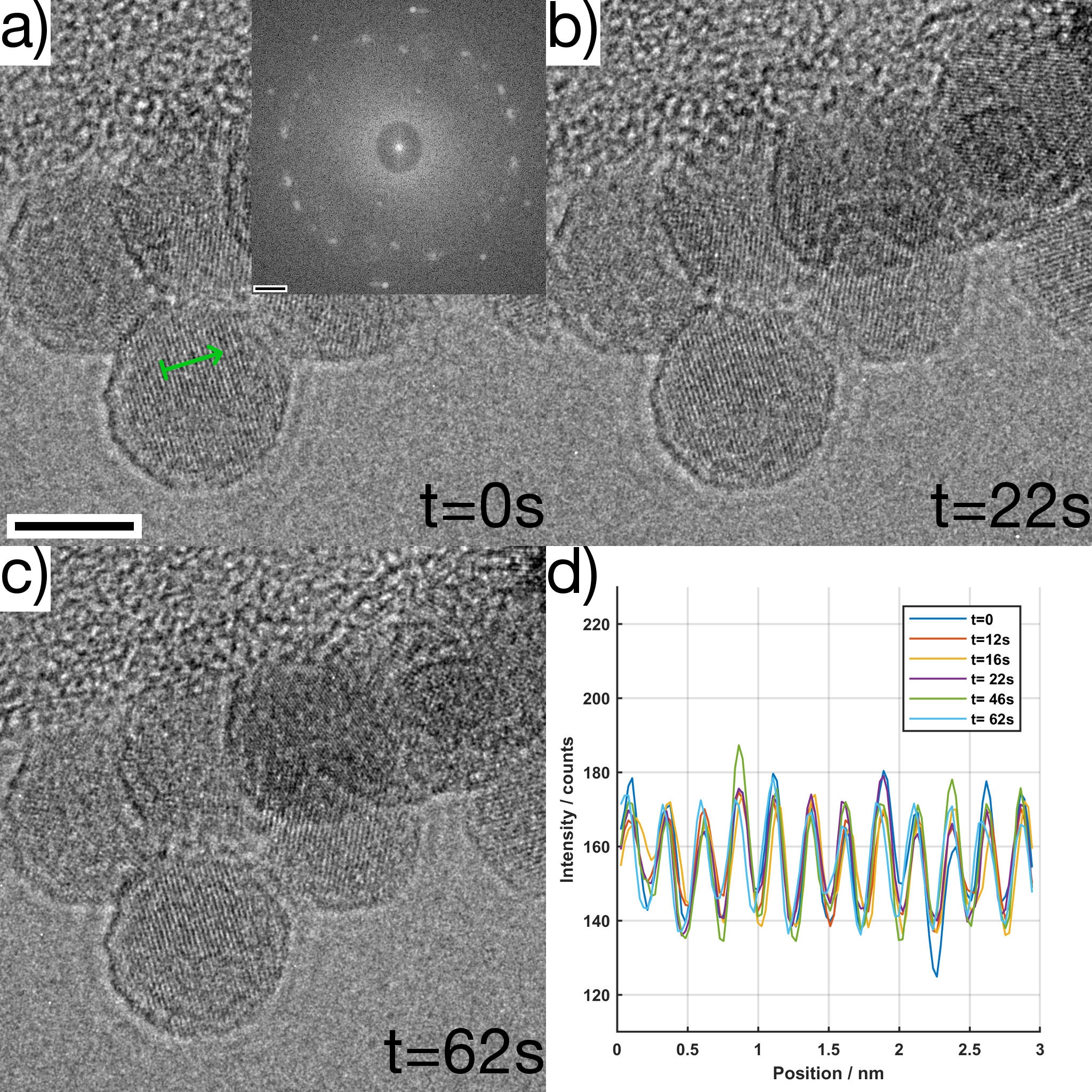}
    \caption{(a-c) Three high-resolution aberration-corrected TEM images of Fe\textsubscript{3-$\updelta$}O\textsubscript{4} nanoparticles acquired with Zernike phase plate (3.5 $\upmu$m hole radius)  without changing any parameter. Time passed after initial image is indicated. Image appearance is stable for over one minute, which is proven by the (d) line profiles taken along the arrow indicated in (a). Scale bar is 6 nm and 1 nm\textsuperscript{-1} for inset power spectrum in (a). }
    \label{F:Stab}
\end{figure}

\newpage

\begin{figure}[h]
\centering
    \includegraphics[width=0.9\linewidth]{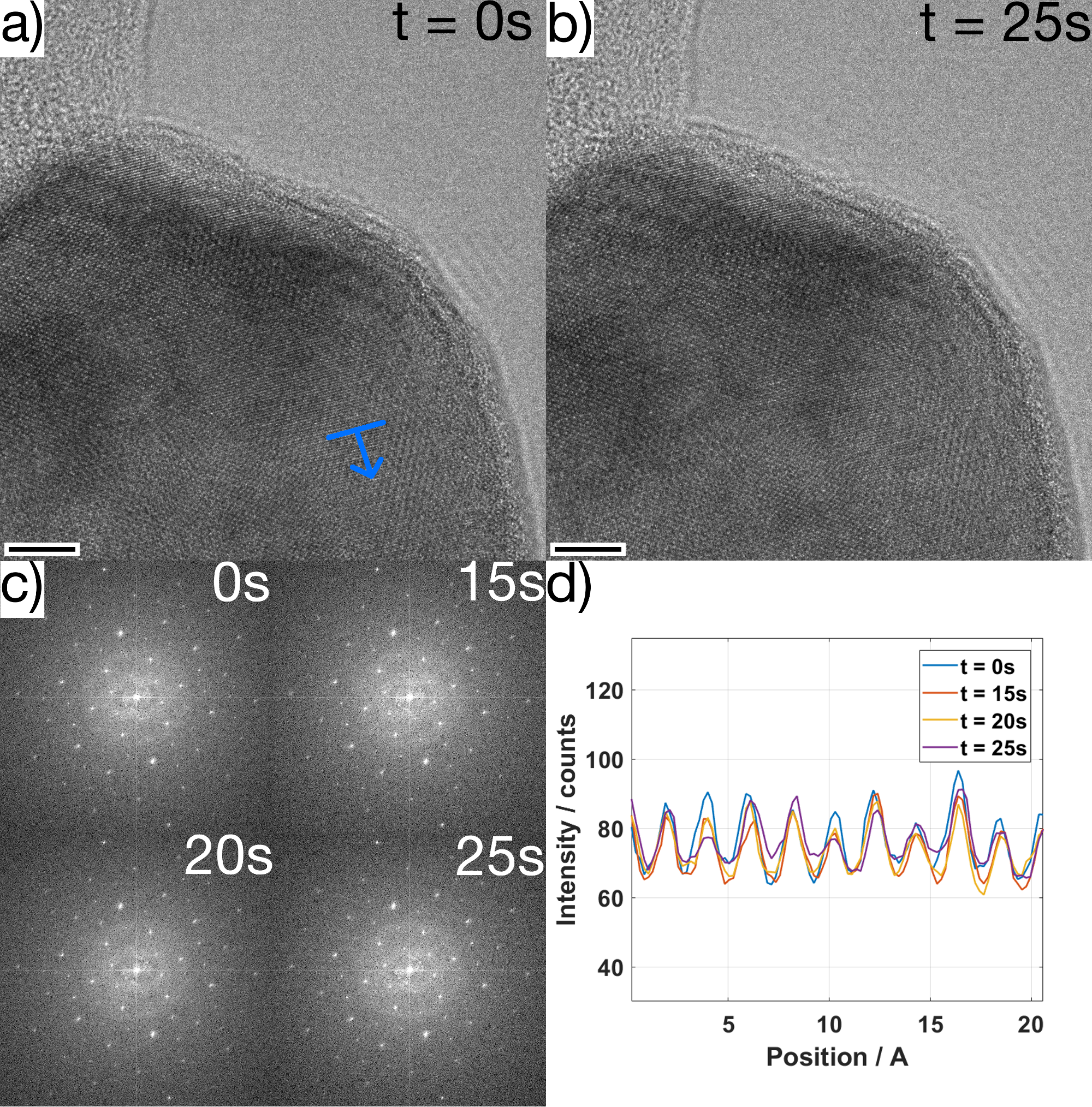}
    \caption{(a,b) Two high-resolution aberration-corrected TEM images of a MoS\textsubscript{2} nanoparticle acquired with Zernike phase plate (4 $\upmu$m hole radius) without changing any parameter. (c) Comparison of power spectra reveals similar appearance. (d) Line profiles taken along the arrow indicated in (a) shows similar contrast and indicates stable imaging conditions. Scale bar is 5 nm and power spectra have a size of 20 nm\textsuperscript{-1} x 20~nm\textsuperscript{-1}. }
    \label{F:StabMos2}
\end{figure}

\newpage

\begin{figure}[h]
\centering
    \includegraphics[width=0.9\linewidth]{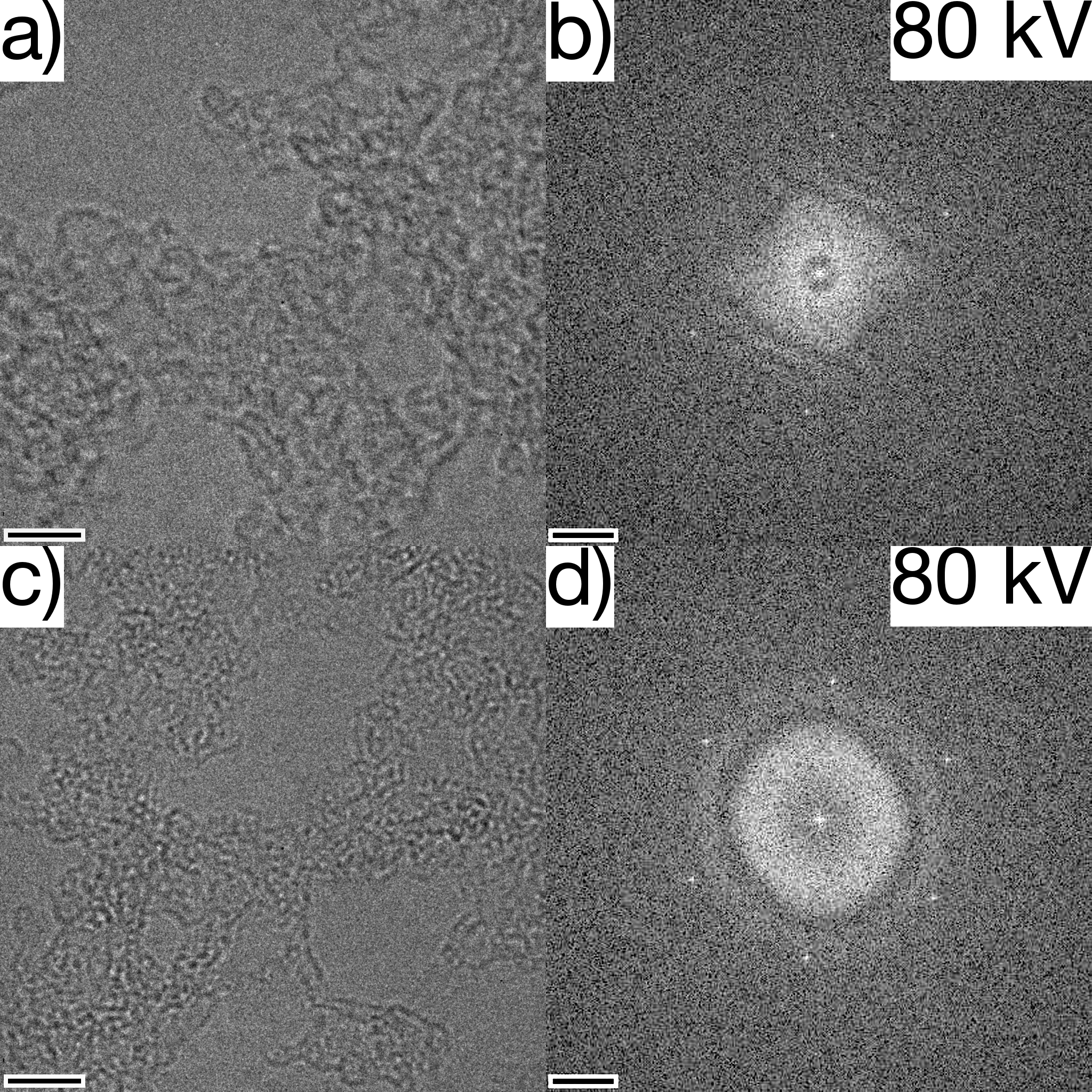}
    \caption{(a,c) Two HRTEM images of a graphene sample acquired at 80~kV (a) with and (c) without a Zernike phase plate (5 $\upmu$m hole radius) and (b,d) corresponding power spectra. The charge build up under the reflections of the graphene manifests itself in a distortion of the image and the corresponding power spectra. Scale bars are (a,c) 3~nm and (b,d) 2~nm\textsuperscript{-1}.}
    \label{F:GrapCharg}
\end{figure}
\newpage

\section{Comparison of HRTEM images with and without PP}

\begin{figure}[h]
\centering
    \includegraphics[width=0.9\linewidth]{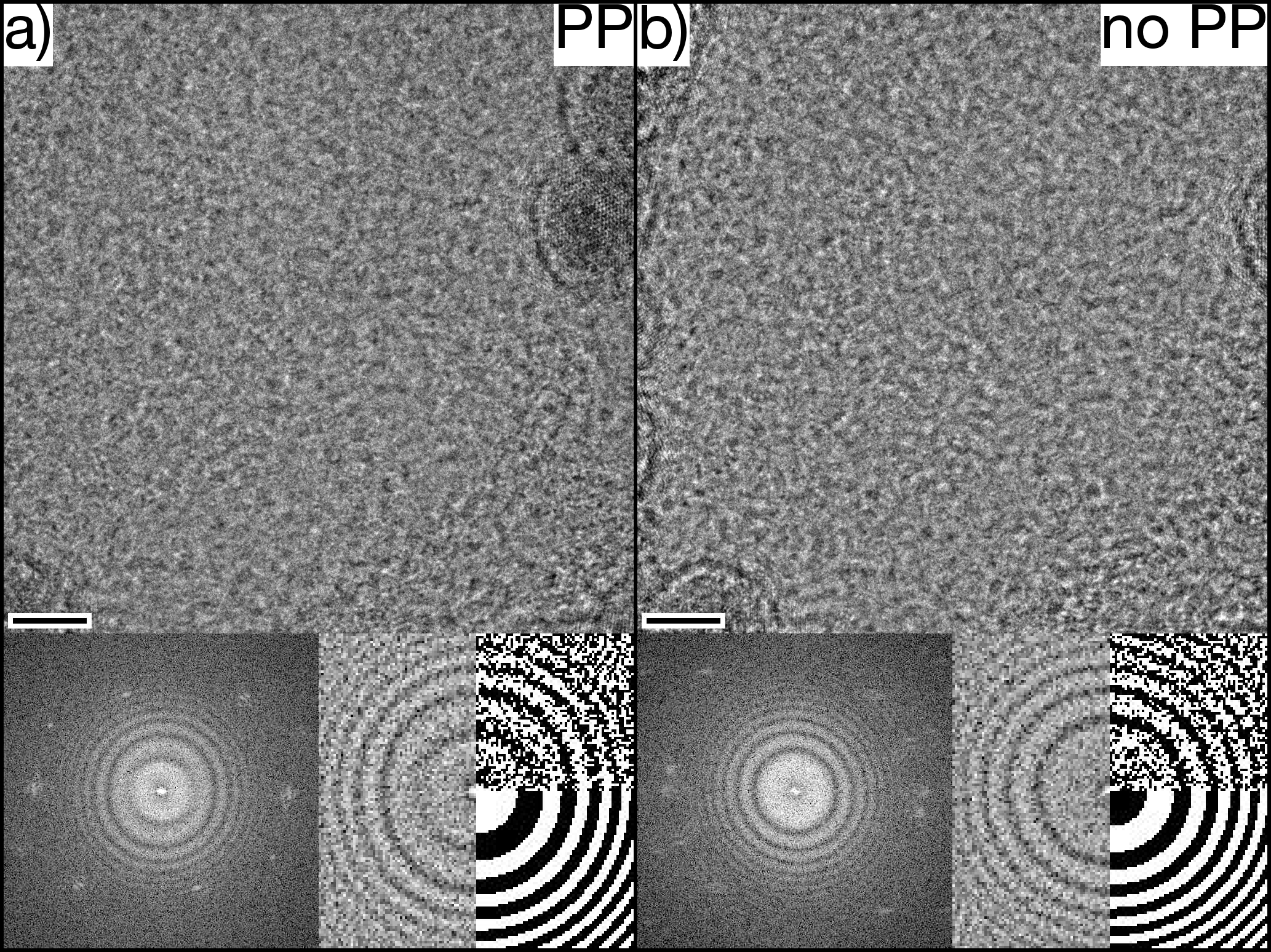}
    \caption{Aberration-corrected TEM image of aC film with Fe\textsubscript{3-$\updelta$}O\textsubscript{4} nanoparticles and corresponding power spectra acquired (a) with and (b) without Zernike phase plate (radius of 4 $\upmu$m) without changing any additional parameter. Defocus is measured to (a) 266 nm and (b) 296 nm, phase shift is (a) 0.45$\pi$. Mean image intensity is (a) 197.8 and (b) 200.3. Scale bars are 5~nm and power spectra/fitted spectra are displayed up to 5 nm\textsuperscript{-1} and 2.86 nm\textsuperscript{-1}, respectively. }
    \label{F:PPnoPP}
\end{figure}
\newpage

\begin{figure}[h]
\centering
    \includegraphics[width=0.9\linewidth]{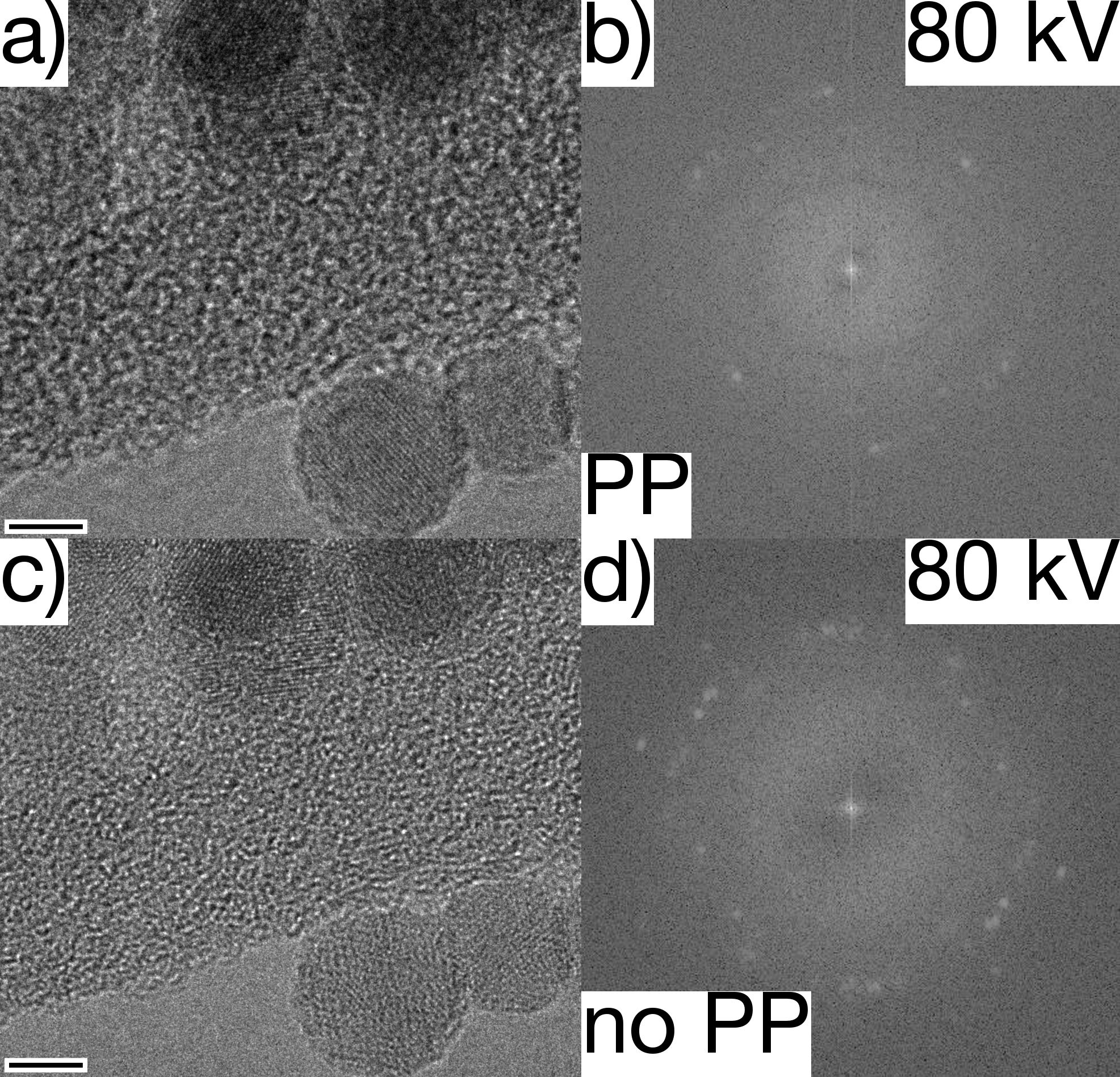}
    \caption{Aberration-corrected TEM image of identical sample region of an aC film with Fe\textsubscript{3-$\updelta$}O\textsubscript{4} nanoparticles and corresponding power spectra acquired at 80 kV (a,b) with and (c,d) without Zernike phase plate (radius of 6 $\upmu$m). Mean image intensity is (a) 58.4 and (b) 61.4. Scale bars are 4~nm and power spectra are displayed up to 5.6 nm\textsuperscript{-1}. }
    \label{F:PPnoPP80}
\end{figure}

\section*{Acknowledgements}

The authors acknowledge funding by German Research Foundation (DFG project He 7675/1-1), from the European Union’s Horizon 2020 research and innovation programme under the Marie Sklodowska-Curie grant agreement No 889546, by the Spanish MICINN (PID2019-104739GB-100/AEI/10.13039/501100011033) and from the European Union H2020 programs “ESTEEM3” (Grant number 823717) and "Graphene Flagship" CORE 3 (Grant number 881603). The microscopy works have been conducted in the Laboratorio de Microscopias Avanzadas (LMA) at Universidad de Zaragoza. Support from R. Valero (LMA) with electron-beam evaporation of aC thin films is acknowledged. Sample courtesy from K. Sartori and B. Pichon (Fe\textsubscript{3-$\updelta$}O\textsubscript{4} NP, Universite de Strasbourg, CNRS, France), MB Sreedhara and R. Tenne (W(S,Se)\textsubscript{2}, Weizmann Institute of Science, Israel) and A. Ismach (MoS\textsubscript{2}, Tel Aviv University, Israel) is acknowledged.